\documentclass[aps,prl,twocolumn,floats,superscriptaddress,tighten,letterpaper,floatfix]{revtex4}
\usepackage{bbm,mathtools}
\usepackage{amssymb}
\usepackage{amsbsy}
\usepackage{amsmath}
\usepackage{graphicx}
\usepackage{graphics}
\usepackage{setspace}
\usepackage{array}
\usepackage{color}
\usepackage{xcolor}
\usepackage{fontenc}
\usepackage{textcomp}
\usepackage{rotating}
\usepackage{bm}
\usepackage{braket}
\usepackage{hyperref}
\hypersetup{pdfpagemode=UseNone}
\usepackage{chngcntr}
\usepackage{subfigure}

\DeclareMathOperator{\e}{e}
\DeclareMathOperator{\sech}{sech}
\DeclareMathOperator{\diag}{diag}
\newcommand{\vex}[1]{\bm{\mathrm{#1}}}

\newcommand{\bsub}{\begin{subequations}}
\newcommand{\esub}{\end{subequations}}
\newcommand{\sfs}{\mathsf{s}}
\newcommand{\sfb}{\mathsf{b}}
\newcommand{\ord}[1]{\bm{\mathit{O}}\left(#1\right)}
\newcommand{\LL}{\lambda_L^{{\scriptscriptstyle{(0)}}}}
\newcommand{\lcoh}{l_{\mathsf{coh}}}
\newcommand{\lfb}{\lambdabar_F}

\begin{document}

\title{Power-law Temperature Dependence of the Penetration Depth in a Topological Superconductor due to Surface States}

\author{Tsz Chun Wu}
\affiliation{Department of Physics and Astronomy, Rice University, Houston, Texas 77005, USA}
\author{Hridis K. Pal}
\affiliation{Department of Physics, Indian Institute of Technology Bombay, Powai, Mumbai 400076, India}
\author{Pavan Hosur}
\affiliation{Texas Center for Superconductivity and Department of Physics, University of Houston, Houston, Texas 77204, USA}
\author{Matthew S. Foster}
\affiliation{Department of Physics and Astronomy, Rice University, Houston, Texas 77005, USA}
\affiliation{Rice Center for Quantum Materials, Rice University, Houston, Texas 77005, USA}

\date{\today}

\begin{abstract}
We study the temperature dependence of the magnetic penetration depth in a 3D topological superconductor (TSC), 
incorporating the paramagnetic current due to the surface states. A TSC is predicted to host a gapless 2D surface 
Majorana fluid. In addition to the bulk-dominated London response, we identify a $T^3$ power-law-in-temperature 
contribution from the surface, valid in the low-temperature limit. Our system is fully gapped in the bulk, and 
should be compared to bulk nodal superconductivity, which also exhibits power-law behavior. Power-law temperature 
dependence of the penetration depth can be one indicator of topological superconductivity.
\end{abstract}

\maketitle

A decade after the widespread infiltration of topology into quantum materials research, 
the search for electronically correlated topological phases beyond the fractional quantum Hall effect
remains an urgent, but still largely unfulfilled quest. Topological superconductivity 
\cite{Qi_Zhang_Review,Sato_Ando_Review}
is sought 
as a platform for Majorana fermion zero modes 
\cite{Jason_Review}
and 
topological quantum computation 
\cite{Q_computer1}. 
Majorana fermions could be detected 
by various means, including tunneling spectroscopy
\cite{MF_STM1,MF_STM2,MF_STM3,MF_STM4,MF_STM5},
the Josephson effect 
\cite{4pi_JE1,4pi_JE2,MF_STM1},
as well as spin and optical responses \cite{spin_resp_TSC,optical_resp_TSC}. 

Only a handful of materials have emerged as bulk topological
superconductor (TSC) candidates.
Topological superconductivity has long been suspected in Sr$_2$RuO$_4$, 
although a consensus on chiral $p$-wave order is yet to be reached 
\cite{Sato_Ando_Review,Mackenzie-Maeno03,Kallin12,Kallin16}.
Time-reversal invariant TSCs 
could serve as solid-state
analogs of the topological superfluid phase in liquid helium 
($^3$He-$B$) \cite{Ludwig08,Roy08,Volovik09,Qi09,Sato09,Sato10,Machida16}.
In the absence of a magnetic field, the predicted hallmark of such a bulk TSC 
is a gapless, two-dimensional (2D) Majorana fermion surface fluid. 
It has been argued that the odd-parity 
``$\hat{\vex{s}}\cdot\vex{k}$'' 
pairing of $^3$He-$B$ 
\cite{VolovikBook,BernevigHughesBook}
could naturally arise 
in doped Dirac semimetals or topological insulators \cite{CuBiSe_pairing,CuBiSe_Anisopairing,Wray10};
here $\hat{\vex{s}}$ and $\vex{k}$ respectively denote the spin-operator and momentum vectors. 
Alternately, doped Weyl semimetals have been shown to be natural platforms for topological superconductivity \cite{Hosur14}. 

There is now substantial experimental evidence for nematicity in 
the 
superconducting
doped topological insulators (Cu,Nb)$_x$Bi$_2$Se$_3$ 
\cite{Zheng16,Yonezawa17,Qiu17,Tao18} (see also \cite{Wen18}), 
possibly indicative of odd-parity pairing.
However, gapless Majorana fermions have not been conclusively detected
\cite{Ando11,Chu13,Stroscio13,Tao18}.
Recently, signatures consistent with topological superconductivity were also found in doped $\beta$-PdBi$_2$, 
but a conclusive detection remains elusive \cite{Kolapo18}.
In Cu$_x$Bi$_2$Se$_3$, 
only a small percentage of the exposed crystal surface was found to exhibit signatures of superconductivity
in STM \cite{Tao18}, highlighting the possibility that 
in inhomogeneous TSCs, there is no guarantee that Majorana fermions 
will appear at the \emph{physical} surface of the sample. 

It is therefore natural to seek global probes of topological superconductivity. 
Meissner effect penetration depth measurements in 
Nb$_x$Bi$_2$Se$_3$ \cite{Smylie16,Smylie17} 
and in
the half-Heusler compounds YPtBi \cite{Paglione18},
YPdBi and TbPdBi \cite{Spinu18} 
exhibit power-law temperature suppression,
which is interpreted as evidence for non-$s$-wave, 
bulk nodal
superconductivity \cite{Tinkham}. 
In the case of Nb$_x$Bi$_2$Se$_3$, the results were interpreted as 
indicative of nodal odd-parity bulk pairing \cite{Smylie16,Smylie17},
while the YPtBi results were attributed to either an exotic nodal-line
``septet'' pairing scenario \cite{Paglione18,Brydon16}, 
or $d + i d$ bulk Weyl superconductivity \cite{Brydon16}, smeared by disorder \cite{Roy19}.

%%%%%%%%%%%%%%%%%%%%%%%%%%%%%%%%%%%%%%%%%%%%%%%%%%%%%%%%%%%
\begin{figure}[b]
\includegraphics[width=0.45\textwidth]{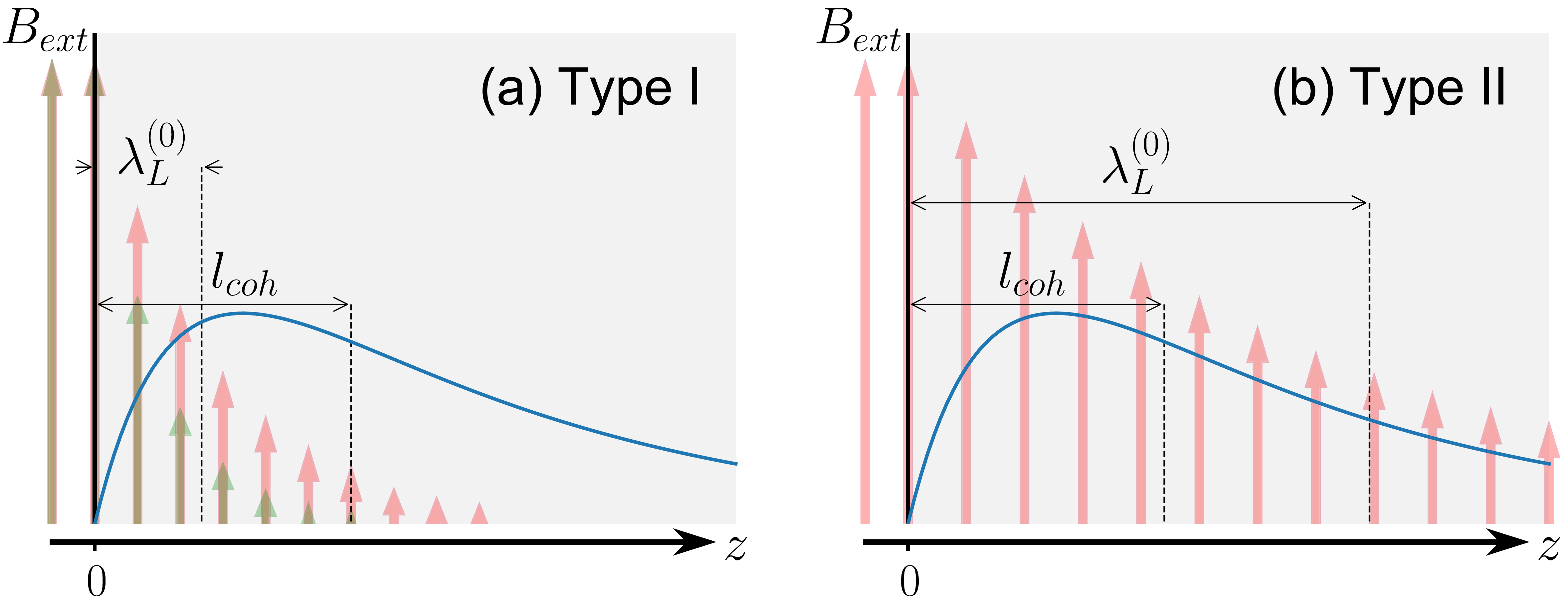}
\caption{Schematic illustration for (a) type I and (b) type II topological superconductors (TSCs)
in an external magnetic field. The gray and white regions represent respectively the TSC
%topological superconductor 
and the vacuum. 
The red (green) arrows indicate the magnetic field that decays into 
the superconductor with (without) the correction due to the surface states.
The scale $\LL$ is the London (bulk-dominated diamagnetic) penetration depth. 
The blue line is a sketch of the Majorana surface fluid density, characterized 
by the coherence length $l_{\mathsf{coh}}$.}	
\label{fig:geometry}
\end{figure}
%%%%%%%%%%%%%%%%%%%%%%%%%%%%%%%%%%%%%%%%%%%%%%%%%%%%%%%%%%%

In this Letter, we show that power-law temperature ($T$) dependence of the penetration
depth $\lambda(T)$ can arise at arbitrarily low $T$ in a TSC with a fully gapped bulk, 
due to the paramagnetic response of the gapless Majorana surface fluid. 
This requires a nontrivial calculation employing a TSC model with a physical surface-vacuum boundary, 
and the result involves the convolution of the surface paramagnetic and bulk-dominated 
diamagnetic responses. It cannot be obtained from the surface Hamiltonian alone. 
Thus, the observation of non-exponential behavior in $\lambda(T)$ does not 
necessarily indicate bulk nodal pairing, and can be one diagnostic for 
screening possible TSCs. 
We also show that the magnetic field in a fully gapped TSC is sensitive to the spatial profile of the 
surface states. This can be used to differentiate gapless states from the surface versus those from the bulk. 
By contrast, STM is sensitive only to the sample surface density of states, while the finite energy resolution
of ARPES might prevent the direct detection of Majorana fermions in a low-temperature superconductor. 
While power-law temperature dependence in the penetration depth can arise from multiple mechanisms \cite{Giannetta06,Cooper97}, 
it can serve as one possible indicator for bulk topological superconductivity.
In general, one wants as many independent tests as possible in order to identify TSCs.

We consider ``minimal'' TSCs in class DIII with winding number $\nu = 1$,
possessing a single surface Majorana cone. 
We show that the leading correction to the London response 
due to the presence of Majorana surface states scales as $T^3$. 
The same temperature dependence is predicted to arise
in the suppression of the mass flow of superfluid $^3$He-$B$
through a channel, due to surface currents \cite{Wu13}.
We calculate explicitly the magnetic field profile inside the slab, which incorporates new features introduced by the surface states. 
For type I TSCs, the field penetrates much deeper than the London depth into the bulk, with the scale set by the coherence length. 
For type II TSCs the field is modulated in a shallow region near the surface, and
then decays at deeper depths according to the London length, 
but with an enhanced field amplitude.

%%%%%%%%%%%%%%%%%%%%%%%%%%%%%%%%%%%%%%%%%%%%%%%%%%%%%%%%%%%%%%%%%%%%%%%%%%%%%%%%%
%%%%%%%%%%%%%%%%%%%%%%%%%%%%%%%%%%%%%%%%%%%%%%%%%%%%%%%%%%%%%%%%%%%%%%%%%%%%%%%%%
%%%%%%%%%%%%%%%%%%%%%%%%%%%%%%%%%%%%%%%%%%%%%%%%%%%%%%%%%%%%%%%%%%%%%%%%%%%%%%%%%
%%%%%%%%%%%%%%%%%%%%%%%%%%%%%%%%%%%%%%%%%%%%%%%%%%%%%%%%%%%%%%%%%%%%%%%%%%%%%%%%%
%%%%%%%%%%%%%%%%%%%%%%%%%%%%%%%%%%%%%%%%%%%%%%%%%%%%%%%%%%%%%%%%%%%%%%%%%%%%%%%%%
%%%%%%%%%%%%%%%%%%%%%%%%%%%%%%%%%%%%%%%%%%%%%%%%%%%%%%%%%%%%%%%%%%%%%%%%%%%%%%%%%
%%%%%%%%%%%%%%%%%%%%%%%%%%%%%%%%%%%%%%%%%%%%%%%%%%%%%%%%%%%%%%%%%%%%%%%%%%%%%%%%%
%%%%%%%%%%%%%%%%%%%%%%%%%%%%%%%%%%%%%%%%%%%%%%%%%%%%%%%%%%%%%%%%%%%%%%%%%%%%%%%%%
%%%%%%%%%%%%%%%%%%%%%%%%%%%%%%%%%%%%%%%%%%%%%%%%%%%%%%%%%%%%%%%%%%%%%%%%%%%%%%%%%
%%%%%%%%%%%%%%%%%%%%%%%%%%%%%%%%%%%%%%%%%%%%%%%%%%%%%%%%%%%%%%%%%%%%%%%%%%%%%%%%%
%%%%%%%%%%%%%%%%%%%%%%%%%%%%%%%%%%%%%%%%%%%%%%%%%%%%%%%%%%%%%%%%%%%%%%%%%%%%%%%%%

\textit{Model}.---We consider a superconducting slab filling up the $z>0$ half-space, 
with an external magnetic field 
$\textbf{B}_{\mathsf{ext}} = B_0 \, \hat{y}$ 
as shown in Fig.~\ref{fig:geometry}.
We assume the field is weak enough such that $B_0 < B_{c_1}$ where $B_{c_1}$ is the lower critical field of a type II superconductor. Under this assumption, the effect of vortex formation can be neglected and a linear response treatment is valid. The total current $\vex{J}$ and the vector potential $\vex{A}$ satisfy the static Maxwell's equation
\begin{equation}
	\bm{\nabla}^2 \textbf{A}(z) 
	= 
	-\frac{4\pi}{c}\textbf{J}(z),
\end{equation}   
where 
$\textbf{A}(z) = A(z) \, \hat{x}$, 
$\vex{B}(z) = \partial_z A(z) \, \hat{y}$,
$\textbf{J}(z) = J(z) \, \hat{x}$, 
and
\begin{equation}
	\textbf{J}(z) 
	= 
	\textbf{J}_{\mathsf{ext}}(z)
	-
	\frac{1}{c}
	\int_{0}^{\infty}dz'\, \Pi(z,z') \, \textbf{A}(z').
\end{equation}
Here, $\textbf{J}_{\mathsf{ext}}(z)$ is a fictitious current generating the external magnetic field, 
and 
$\Pi(z,z')$ is the current-current correlation function capturing the linear response of the TSC. The above equations can be expressed as
\begin{equation}\label{eq:IntegralEqForB}
\begin{aligned}[b]
	- 
	2 B_0 
	=&\, 
	\left[
		q_z^2 + \left(\LL\right)^{-2}
	\right]
	\tilde{A}(q_z)
\\
	&\,
	+
	\frac{4\pi}{c^2}
	\int
	\frac{d Q_z}{2 \pi}
	\,
	\tilde{\Pi}^{xx}_{1,R}(0,0,0;q_z,-Q_z)
	\,
	\tilde{A}(Q_z),
\end{aligned}
\end{equation}
where 
$\LL$
is the London penetration depth and 
\begin{equation}
\begin{aligned}[b]
	&\tilde{\Pi}^{xx}_{1,R}(\Omega = 0,q_x = 0,q_y = 0;q_z,-Q_z) \\
	&\;\;= 
	-i 
	\int dt \, dx \, dy 
	\int_0^{\infty}dz_1 dz_2 
	\e^{-i q_z z_1 + i Q_z z_2}\\
	&\qquad\qquad\qquad
	\times
	\left\langle \left[J_1^{x}(t,x,y,z_1),J_1^{x}(0,0,0,z_2)\right]\right\rangle \theta(t)
\end{aligned}
\end{equation}
is the retarded paramagnetic current-current correlation function. 
Here, $J_1^x(t,x,y,z)$ is the paramagnetic current flowing along the $x$ direction and $\theta(t)$ is the Heaviside step function. 
The first term 
on the right-hand-side
of Eq.~(\ref{eq:IntegralEqForB}) represents the diamagnetic London response,
while the second term is the paramagnetic response from \textit{both} the bulk and surface states. 
The above framework is general and the magnetic field in the slab is determined once the current-current correlation function is specified. 
In what follows we consider a clean system. Weak nonmagnetic disorder is not expected to modify the low-temperature response of the 
surface response (it is strongly irrelevant \cite{Ghorashi19}) or of the fully gapped bulk. 

Although the low-temperature, $T^3$-dependence of the penetration depth derived below depends only on the low-energy
dispersion of the 2D Majorana surface fluid, here we consider a microscopic model for both the bulk and surface modes
of the TSC in order to completely specify the problem. 
Solid state models analogous to $^3$He provide a fertile playground to study topologically nontrivial superconductivity 
\cite{Sato_Ando_Review,VolovikBook,Volovik09}.
``Solid-state $^3$He-$A$'' would correspond to a Weyl superconductor, 
which has nodal Weyl points in the bulk that connect to a surface Majorana arc. 
In this work, we consider ``solid-state $^3$He-$B$'' \cite{VolovikBook,BernevigHughesBook}, 
with isotropic $p$-wave pairing of spin-1/2 electrons, 
represented by the following  
Bogoliubov-de Gennes
Hamiltonian 
\begin{equation}\label{HDef}
	H 
	= 
	\frac{1}{2}
	\int_\textbf{k} 
	\chi^{\dagger}(\textbf{k}) 
	\,
	\hat{h}(\textbf{k})
	\,
	\chi(\textbf{k}), 
\end{equation}
where
\begin{equation}\label{hDef}
	\hat{h}(\textbf{k}) = \tilde{\varepsilon}_k \, \hat{\sigma}^3 + \Delta \, \hat{\textbf{s}}\cdot \textbf{k} \, \hat{\sigma}^2,
\qquad
	\tilde{\varepsilon}_k = \frac{k^2}{2m} - \mu,
\end{equation}
and where 
$
	\int_\textbf{k} \equiv \int \frac{d^3\textbf{k}}{(2\pi)^3},
$
$\hat{\textbf{s}}$ and $\bm{\hat{\sigma}}$ respectively denote Pauli matrices 
acting in the spin and particle-hole spaces, 
and 
$
	\chi(\textbf{k}) \equiv \left[c(\textbf{k}), \hat{s}^2\left[c^{\dagger}(-\textbf{k})\right]^{\mathsf{T}}\right]
$ 
is the four-component Balian-Werthammer spinor \cite{BW63}. 
The latter satisfies the reality (``Majorana'') condition 
$
	\chi^{\dagger}(\textbf{k})= i\chi^{\mathsf{T}}(-\textbf{k}) \, \hat{M}_{\mathsf{P}},
$ 
where $\hat{M}_{\mathsf{P}} = \hat{s}^2 \hat{\sigma}^2$ defines particle-hole symmetry for $\hat{h}(\vex{k})$. 
In Eq.~(\ref{hDef}),
$\mu$ is the chemical potential and $\Delta$ is the superconducting order parameter amplitude. For $\mu > 0$ the above Hamiltonian has winding number $\nu = 1$. The retarded paramagnetic current-current correlation function due to the bulk is 
\begin{equation}\label{paraBulk}
	\!\!\!
	\tilde{\Pi}_{1,R}^{xx}(\Omega = 0,\textbf{q}=0) 
	= 
	-
	\frac{\beta}{6}\left(\frac{e}{m}\right)^2 
	\int_\textbf{k} k^2 \sech^2\left(\frac{\beta E_k}{2}\right)\!,\!\!\!
\end{equation}
where $E_k = \sqrt{\tilde{\varepsilon}_k^2 + \Delta^2 k^2}$ is the eigenenergy of $\hat{h}(\textbf{k})$
and $\beta = T^{-1}$ is the inverse temperature.  
In the low-temperature limit, the bulk paramagnetic response is exponentially suppressed 
and the diamagnetic London response dominates. 
The London depth is given by 
$
	\LL = \sqrt{mc^2 / (4\pi e^2 n)},
$ 
where $n$ is the charge number density. 

To consider the response from the surface, we 
replace 
$
	k_z \rightarrow -i\partial_z
$ 
in 
$\hat{h}(\textbf{k})$ [Eq.~(\ref{hDef})]
and solve for the Majorana surface states 
$\psi_\textbf{k}^{\mathsf{s}}(z)$ with eigenenergies $\pm \Delta |\textbf{k}|$.
Here and in what follows, $\textbf{k}= (k_x,k_y)$ specifies the momentum transverse to the interface.
With hard wall boundary conditions at $z = 0$, we obtain surface wave functions of the form
\cite{SM}
\begin{equation}\label{psiSurf}
	\psi_\textbf{k}^{\mathsf{s}}(z) 
	= 
	\frac{e^{-z / l_{\mathsf{coh}}}}{\sqrt{N_\textbf{k}^{\mathsf{s}}}}
	\,
	\sin\left(z\sqrt{\lambdabar_F^{-2}-l_{\mathsf{coh}}^{-2}-k^2} \right)
	\ket{\psi_\textbf{k}^{0}},
\end{equation} 
where $N_k^{\mathsf{s}}$ is a normalization constant and $\ket{\psi_\textbf{k}^{0}}$ is a 
spin-momentum-locked
spinor in (spin)$\otimes$(particle-hole) space.  
The two length scales in Eq.~(\ref{psiSurf}) are
the (reduced) Fermi wavelength 
$\lambdabar_F \equiv 1/\sqrt{2 m \mu}$
and the coherence length 
$l_{\mathsf{coh}} \equiv 1/m\Delta$. 

We can see how the magnetic field couples to the surface fluid
by incorporating a vector potential $\vex{A}$ into Eq.~(\ref{HDef}),
and then projecting onto the low-energy surface states. 
The result is 
\begin{equation}\label{HsDef}
	H_{\mathsf{s}} = \int_\textbf{r} 
	\left\lbrace
	\frac{1}{2}\eta^{\dagger}(\textbf{r})
	\left[	
		-i\Delta(\hat{s}\wedge \bm{\nabla})
	\right]
	\eta(\textbf{r})
	- 
	\frac{1}{c}\textbf{A}\cdot \textbf{J}
	\right\rbrace, 
\end{equation}
where $\eta = \eta_{s}$ is the two-component surface Majorana fermion operator
($s \in \{\uparrow,\downarrow\}$), 
$	
	\hat{s}\wedge \bm{\nabla} \equiv \hat{s}^1 \partial_y - \hat{s}^2 \partial_x,
$ 
and the surface paramagnetic current operator
\begin{equation}\label{SurfCurr}
	\textbf{J}(\textbf{r}) 
	= 
	\frac{e}{4m}
	\eta^{\dagger}(\textbf{r}) 
	i\bm{\overleftrightarrow{\nabla}}
	\eta(\textbf{r}).
\end{equation}
Here $\textbf{r} = (x,y)$ and $\bm{\overleftrightarrow{\nabla}} \equiv \bm{\overrightarrow{\nabla}}-\bm{\overleftarrow{\nabla}}$. 
Eq.~(\ref{HsDef}) assumes that the field $\vex{B}(z)$ and the vector potential $\vex{A}(z)$ (in London gauge) 
both reside in the $(x,y)$ plane. 
Zeeman coupling to a nonzero component $B_z$ would induce a Majorana mass,
gapping out the surface fluid \cite{Qi_Zhang_Review}, 
but this is prevented by bulk Meissner screening. 
On the other hand, a very strong in-plane field $A_{x,y}$ could ``overtilt''
the surface Majorana cone, creating a surface Fermi pocket. The latter should be included in the 
diamagnetic current \cite{Guinea19}, but we exclude this situation here by restricting 
to linear response.  

In the low-temperature limit, 
the paramagnetic current-current correlation function due to the surface state fluid 
evaluates to \cite{SM} 
\begin{equation}\label{paraSS}
\begin{aligned}[b]
	&\tilde{\Pi}_{1,R,\sfs,\sfs}^{xx}(0,0,0;q_z,-Q_z) \\
	&\qquad
	\simeq
	-
	\mathcal{C}
	\,
	\left(\frac{e}{m}\right)^2 
	\frac{(k_B T)^3}{\Delta^4}
	\,
	\Theta(0,q_z)
	\,
	\Theta(0,-Q_z),
\end{aligned}
\end{equation}
$\mathcal{C} \equiv \left[3^2 \zeta(3)/(2^3 \pi)\right]$,
$\zeta(z)$ is the Riemann Zeta function,
and where
\begin{equation}
	\Theta(k,q_z)
	\equiv
	\int_0^{\infty}dz\,
	\e^{-iq_z z}
	\,
	\psi_\textbf{k}^{\mathsf{s}\dagger}(z)
	\,
	\psi_\textbf{k}^{\mathsf{s}}(z)
\end{equation}
is the Fourier transformed probability density of the surface states along the $z$-direction.
Unlike the paramagnetic response from the bulk, the one from the surface 
[Eq.~(\ref{paraSS})] 
has a non-trivial $T^3$ power-law dependence at low temperature. 
Two factors of temperature arise from the form of the paramagnetic current operator (a derivative) in Eq.~(\ref{SurfCurr}), 
while the third stems from the surface density of states of the Majorana fluid. 
One should also consider the surface-bulk cross terms when evaluating the paramagnetic current-current correlator
appearing in Eq.~(\ref{eq:IntegralEqForB}).
However, these cross terms exhibit higher-power temperature-dependence at low $T$,
and are thus subleading \cite{SM}. We neglect these surface-bulk contributions in the following.

%%%%%%%%%%%%%%%%%%%%%%%%%%%%%%%%%%%%%%%%%%%%%%%%%%%%%%%%%%%%%%%%%%%%%%%%%%%%%%%%%
%%%%%%%%%%%%%%%%%%%%%%%%%%%%%%%%%%%%%%%%%%%%%%%%%%%%%%%%%%%%%%%%%%%%%%%%%%%%%%%%%
%%%%%%%%%%%%%%%%%%%%%%%%%%%%%%%%%%%%%%%%%%%%%%%%%%%%%%%%%%%%%%%%%%%%%%%%%%%%%%%%%
%%%%%%%%%%%%%%%%%%%%%%%%%%%%%%%%%%%%%%%%%%%%%%%%%%%%%%%%%%%%%%%%%%%%%%%%%%%%%%%%%
%%%%%%%%%%%%%%%%%%%%%%%%%%%%%%%%%%%%%%%%%%%%%%%%%%%%%%%%%%%%%%%%%%%%%%%%%%%%%%%%%
%%%%%%%%%%%%%%%%%%%%%%%%%%%%%%%%%%%%%%%%%%%%%%%%%%%%%%%%%%%%%%%%%%%%%%%%%%%%%%%%%
%%%%%%%%%%%%%%%%%%%%%%%%%%%%%%%%%%%%%%%%%%%%%%%%%%%%%%%%%%%%%%%%%%%%%%%%%%%%%%%%%
%%%%%%%%%%%%%%%%%%%%%%%%%%%%%%%%%%%%%%%%%%%%%%%%%%%%%%%%%%%%%%%%%%%%%%%%%%%%%%%%%
%%%%%%%%%%%%%%%%%%%%%%%%%%%%%%%%%%%%%%%%%%%%%%%%%%%%%%%%%%%%%%%%%%%%%%%%%%%%%%%%%
%%%%%%%%%%%%%%%%%%%%%%%%%%%%%%%%%%%%%%%%%%%%%%%%%%%%%%%%%%%%%%%%%%%%%%%%%%%%%%%%%
%%%%%%%%%%%%%%%%%%%%%%%%%%%%%%%%%%%%%%%%%%%%%%%%%%%%%%%%%%%%%%%%%%%%%%%%%%%%%%%%%

\textit{Results}.---Taking only the diamagnetic and surface paramagnetic responses into account, 
which is valid at low temperature as discussed above, we can formally invert the integral equation 
Eq.~(\ref{eq:IntegralEqForB}) and solve for the vector potential (and hence the magnetic field) profile inside the slab. 
To leading order in temperature, the final result is \cite{SM}
\begin{equation}\label{eq:B}
	B_y(z) 
	=
	B_0
	\left\{
	\e^{-z/\LL} 
	-
	\varrho(T) \bigg[ \partial_z G(z)\bigg] G(0)
	\right\},
\end{equation}
where
\begin{align}
	\varrho(T) 
	\equiv
	\left[2^4 3^3 \zeta(3) \pi \right] 
	\left(\LL\right)^6 
	\,
	l_{\mathsf{coh}}^{-1}
	\,
	\lambdabar_F^{-4}
	\,
	t^3
\end{align}
is a temperature-dependent length, with $t \equiv k_B T/ \Delta k_F$ being the dimensionless temperature.
Here, $\Delta k_F$ is the energy gap of the $p$-wave TSC.
In Eq.~(\ref{eq:B}),
the function $G(z)$ is a temperature-independent, real-valued function 
encoding the convolution of the bulk and surface responses.
It is a dimensionless function only of $z$ and of the three lengths $\{\lambdabar_F,\LL,l_{\mathsf{coh}}\}$. 
$G(z)$ emerges when we invert the integral equation and Fourier transform the quantities back to real space \cite{SM}. 

The first term in Eq.~(\ref{eq:B}) describes the Meissner screening due to the diamagnetic London response, 
while the second term is the correction due to the Majorana surface fluid. 
The correction term depends on the length $\varrho(T)$ that encodes the $T^3$ dependence,
while its spatial dependence is captured by $G(z)$. 
The function $G(z)$ decays exponentially for large $z$; its spatial extent is governed by the 
maximum of $\{\LL,l_{\mathsf{coh}}/2\}$, assuming that $\lambdabar_F$ is the shortest scale. 
This leads to different qualitative type I and II behaviors.  
Nevertheless, to characterize the overall spatial extent of the magnetic field, 
we can define the effective penetration depth of the system via \cite{Tinkham}
\begin{equation}\label{eq:PenDepth}
	\lambda(T) 
	\equiv 
	\frac{1}{B_0}
	\int_0^{\infty} dz\, B_y(z) 
	= 
	\LL + \varrho(T)\left[G(0)\right]^2.
\end{equation}
The second term in Eq.~(\ref{eq:PenDepth}) is the change of the penetration depth due to the surface states. 
This term is always positive, meaning that the magnetic field can penetrate deeper into the slab for any $T > 0$, due 
to the surface Majorana fluid. 
It is instructive to roughly estimate the order of magnitude for such correction. 
For Cu$_x$Bi$_2$Se$_3$, which is in the extreme type II regime, we substitute typical experimental data 
$\lcoh \sim 10$ nm, $\lambdabar_F \sim 1$ nm, and $\LL \sim 1$ $\mu$m \cite{Ando12,Tao18} to obtain 
$\delta \lambda(T) \sim 0.1 \LL t^3$.

The two physical quantities $B_y(z)$ and $\lambda(T)$ we focus on inherit the $T^3$ dependence from the surface current-current correlation function
[Eq.~(\ref{paraSS})]. 
Similar power-law-dependence is observed in bulk nodal superconductors. 
In contrast to those systems, the model we considered is fully gapped in the bulk, and thus the Majorana surface states are responsible for the 
gapless excitations. 

%%%%%%%%%%%%%%%%%%%%%%%%%%%%%%%%%%%%%%%%%%%%%%%%%%%%%%%%%%%
\begin{figure}
\includegraphics[width=0.4\textwidth]{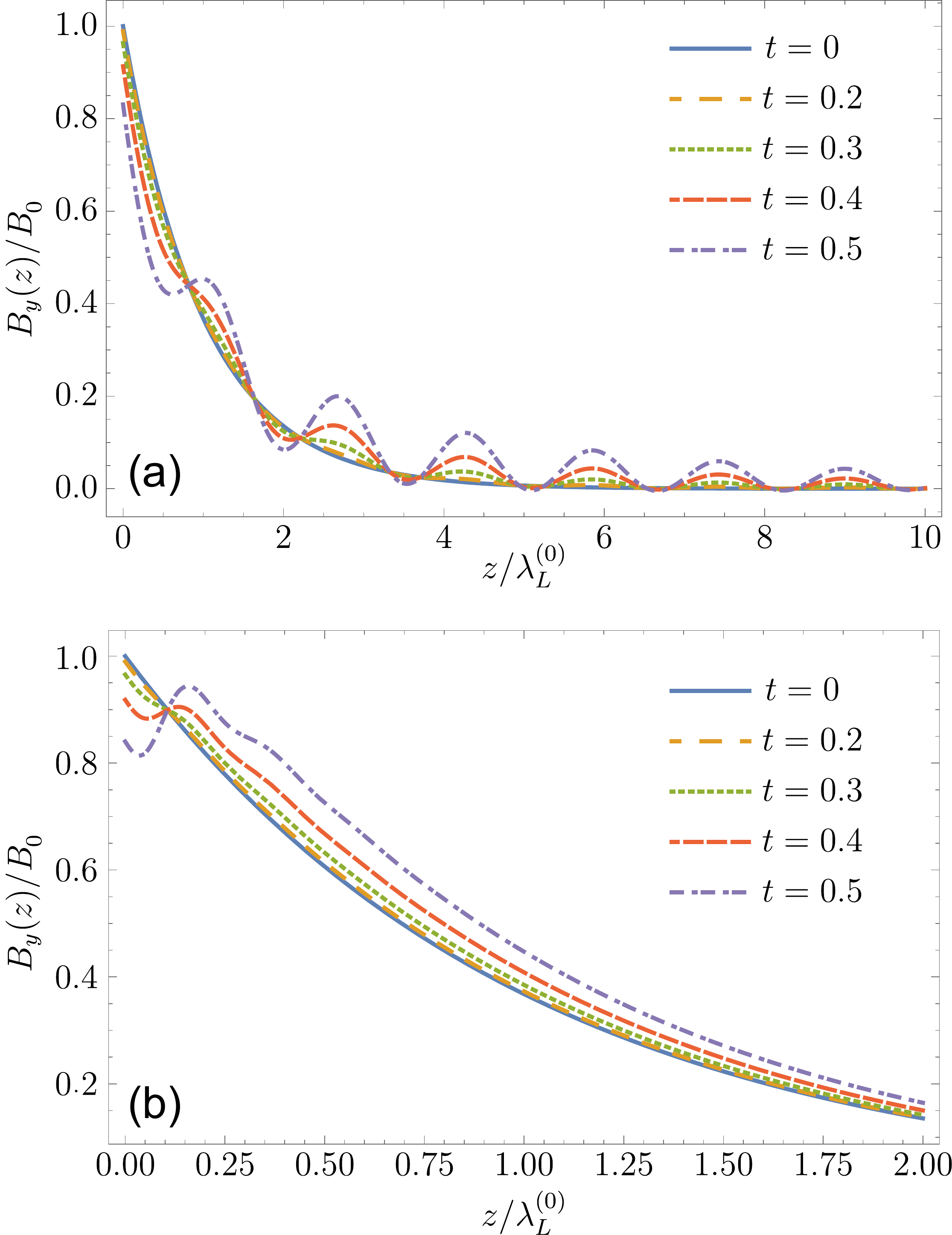}
\caption{
Plot of the magnetic field profile inside the topological superconductor (TSC) at different temperatures in the 
(a) type-I case with $\LL=1$, $l_{\mathsf{coh}}=10$, and $\lambdabar_F = 0.5$, 
and 
(b) type-II case with $\LL=30$, $l_{\mathsf{coh}}=8$, and $\lambdabar_F = 2$. 
Here 
$\lambdabar_F = 1 /\sqrt{2 m \mu}$ is the reduced Fermi wavelength, 
$l_{\mathsf{coh}} = 1 / m \Delta$ is the coherence length, 
and 
$\LL$ is the London depth. 
The blue curves (at $t=0$) represent the diamagnetic London response. 
The dimensionless temperature $t \equiv k_B T / \Delta k_F$, where $\Delta k_F$ is the $p$-wave TSC energy gap. 
As temperature increases, the response from the surface becomes more pronounced. 
For type I (a), the correction from the surface exhibits Friedel oscillations and penetrates much deeper than the London depth.
For type II (b), the strongest modulation due to the surface fluid appears close to the surface.  
}	
\label{fig:B}
\end{figure}
%%%%%%%%%%%%%%%%%%%%%%%%%%%%%%%%%%%%%%%%%%%%%%%%%%%%%%%%%%%

Although the exact expression for $G(z)$ is complicated \cite{SM}, 
it takes relatively simple forms in the strong type-I and type-II limits. 
For a type-I TSC ($l_{\mathsf{coh}} \gg \LL$),
\begin{equation}\label{GI}
	\!\!\!\!\!
\begin{aligned}[b]
	G_{\mathrm{I}}(z) 
	\simeq&\,
	\frac{\lambdabar_F^2}{2\left(\LL\right)^2\left[4\left(\LL\right)^2+\lambdabar_F^2\right]}
	\bigg\{
		-\left(\LL\right)^2 e^{-z/\LL}
\\
	&\,
		+
		\left[	
			2\left(\LL\right)^2 + 
			\lambdabar_F^2 \sin^2
			\left(\frac{z}{\lambdabar_F}\right)
		\right] 
		e^{-2 z/l_{\mathsf{coh}}} 
	\bigg\}.
\end{aligned}
	\!\!\!\!\!
\end{equation}
Since the coherence length sets the depth of the surface fluid [Eq.~(\ref{psiSurf})],
the latter allows a much deeper penetration of the field in the type-I limit than the bulk London depth. 
The slower decay is modulated by a Friedel oscillation. 
Representative field profiles are shown in Fig.~\ref{fig:B}(a).

For a type-II TSC ($l_{\mathsf{coh}} \ll \LL$),
\begin{equation}\label{GII}
	G_{\mathrm{II}}(z) 
	\simeq 
	\frac{\lambdabar_F^2 l_{\mathsf{coh}}}{16\left(\LL\right)^4} 
	\left[
		\LL
		\,
		e^{-z/\LL}
		-
		l_{\mathsf{coh}} 
		\,
		e^{-2z/l_{\mathsf{coh}}} 
	\right].
\end{equation}
In this case the Friedel oscillating terms are subleading, and can be neglected.
Note that the first term in Eq.~(\ref{GII}) dominates the second. 
In this case the spatial field penetration is governed by the London depth,
but the correction in Eqs.~(\ref{eq:B}) and (\ref{GII}) effectively enhances the field amplitude. 
Representative field profiles are indicated in Fig.~\ref{fig:B}(b).

%%%%%%%%%%%%%%%%%%%%%%%%%%%%%%%%%%%%%%%%%%%%%%%%%%%%%%%%%%%%%%%%%%%%%%%%%%%%%%%%%
%%%%%%%%%%%%%%%%%%%%%%%%%%%%%%%%%%%%%%%%%%%%%%%%%%%%%%%%%%%%%%%%%%%%%%%%%%%%%%%%%
%%%%%%%%%%%%%%%%%%%%%%%%%%%%%%%%%%%%%%%%%%%%%%%%%%%%%%%%%%%%%%%%%%%%%%%%%%%%%%%%%
%%%%%%%%%%%%%%%%%%%%%%%%%%%%%%%%%%%%%%%%%%%%%%%%%%%%%%%%%%%%%%%%%%%%%%%%%%%%%%%%%
%%%%%%%%%%%%%%%%%%%%%%%%%%%%%%%%%%%%%%%%%%%%%%%%%%%%%%%%%%%%%%%%%%%%%%%%%%%%%%%%%
%%%%%%%%%%%%%%%%%%%%%%%%%%%%%%%%%%%%%%%%%%%%%%%%%%%%%%%%%%%%%%%%%%%%%%%%%%%%%%%%%
%%%%%%%%%%%%%%%%%%%%%%%%%%%%%%%%%%%%%%%%%%%%%%%%%%%%%%%%%%%%%%%%%%%%%%%%%%%%%%%%%
%%%%%%%%%%%%%%%%%%%%%%%%%%%%%%%%%%%%%%%%%%%%%%%%%%%%%%%%%%%%%%%%%%%%%%%%%%%%%%%%%
%%%%%%%%%%%%%%%%%%%%%%%%%%%%%%%%%%%%%%%%%%%%%%%%%%%%%%%%%%%%%%%%%%%%%%%%%%%%%%%%%
%%%%%%%%%%%%%%%%%%%%%%%%%%%%%%%%%%%%%%%%%%%%%%%%%%%%%%%%%%%%%%%%%%%%%%%%%%%%%%%%%
%%%%%%%%%%%%%%%%%%%%%%%%%%%%%%%%%%%%%%%%%%%%%%%%%%%%%%%%%%%%%%%%%%%%%%%%%%%%%%%%%

\textit{Conclusion}.---Our calculations suggest an alternative way to search for Majorana surface states in TSCs 
by measuring the change in the penetration depth $\delta \lambda(T)$. ARPES is often employed as the key tool 
to detect smoking gun signatures of topology in quantum materials.
Unfortunately, TSC candidates typically have a small gap, 
making surface states difficult for ARPES to resolve \cite{Wray10}. 
A necessary, but not sufficient condition for the existence of a gapless Majorana surface fluid 
can be the signature power-law dependence $\delta \lambda(T) \sim T^3$. 
This can be probed by means of tunnel-diode oscillator techniques \cite{Smylie16,Smylie17,Paglione18,Spinu18}. 
While $\delta \lambda (T)$ is exponentially suppressed in conventional, topologically trivial superconductors, 
a power-law $\delta \lambda(T) \sim T^{\alpha}$ is also expected in superconductors with nodal bulk excitations, 
where $\alpha$ depends on whether the nodes are points or lines~\cite{Tinkham,Sigrist_Review}. 
For example, a power-law with $\alpha = 1$ has been observed in high-$T_c$ $d$-wave superconductors \cite{Giannetta06}. 

One way to distinguish the origin of a penetration-depth power law 
(Majorana surface states versus bulk nodes) 
is via a specific-heat measurement versus temperature. 
The specific heat due to the 2D surface of a fully gapped 3D TSC is negligible. 
In addition to $\delta \lambda(T) \sim T^\alpha$, a TSC with a fully gapped bulk 
(nodal superconductor) should therefore demonstrate exponential suppression 
(power-law dependence) in specific heat \cite{Sigrist_Review}.

The superconducting doped topological insulators (Cu,Nb)$_x$Bi$_2$Se$_3$ 
\cite{Ando12,Smylie16,Smylie17} 
and 
the half-Heusler compound YPtBi \cite{Paglione18,Bay12} 
are all strongly type II.
Power-law dependence of $\lambda(T)$ was observed in 
Nb$_x$Bi$_2$Se$_3$ \cite{Smylie16,Smylie17},
YPtBi \cite{Paglione18},
and in YPdBi and TbPdBi \cite{Spinu18}. 
It would be interesting to assess whether any of these could be attributed to the presence of Majorana surface states. 
In particular, it is worthwhile to note that while specific-heat measurements suggest a fully gapped bulk in Cu$_x$Bi$_2$Se$_3$ and Ni$_x$Bi$_2$Se$_3$\cite{Yonezawa17,Qiu17}, 
penetration depth data for Ni$_x$Bi$_2$Se$_3$ shows a power-law dependence \cite{Smylie16,Smylie17}.
In the case of half-Heusler compounds such as 
YPtBi, it has been suggested that 
optical-phonon-mediated pairing could favor a fully gapped TSC state with winding number $\nu = 3$ 
\cite{Savary17}, and this should induce novel, cubic-dispersing Majorana surface fermions
\cite{Fang15,Roy19}. In this case, we would expect a very slow $\delta \lambda (T) \sim T^{1/3}$ dependence 
for a clean, cubically-dispersing Majorana surface fluid. 
However, the results due to surface states with $\nu \geq 3$ might be strongly modified by quenched
disorder \cite{Roy19,Foster14}. 

In this Letter, calculations were performed for a clean system. 
On one hand, low-energy Majorana surface states with winding number $\nu = 1$ 
are not affected by non-magnetic impurities \cite{Foster14,Ghorashi19}; 
weak nonmagnetic disorder should therefore
not alter the cubic power law $\delta \lambda(T) \sim T^3$ predicted here as $T \rightarrow 0$.  
The power can possibly be modified by strong (resonant) impurity scattering \cite{Giannetta06}.
On the other hand, magnetic impurities can strongly perturb the surface states, 
and even gap them out. Moreover, magnetic impurities may independently induce power-law dependence in $\delta \lambda(T)$,
by altering the magnetic permeability \cite{Giannetta06,Cooper97}. 

We thank 
Andriy Nevidomskyy for useful discussions.
T.C.W.\ and M.S.F.\ acknowledge support 
by the Welch Foundation Grant No.~C-1809, 
by NSF CAREER Grant No.~DMR-1552327,
and by the U.S. Army Research Office
Grant No.~W911NF-17-1-0259.
M.S.F.\ thanks the Aspen Center for Physics, which is
supported by the NSF Grant No.~PHY-1607611, for its
hospitality while part of this work was performed. P.H. and H.K.P. 
were supported by the Department of Physics, 
College of Natural Sciences and Mathematics at the 
University of Houston. 
H.K.P acknowledges support from IRCC, IIT Bombay (RD/0518-IRCCSH0-029).

%%%%%%%%%%%%%%%%%%%%%%%%%%%%%%%%%%%%%%%%%%%%%%%%%%%%%%%%%%%%%%%%%%%%%%%%%%%%%%%%
%%%%%%%%%%%%%%%%%%%%%%%%%%%%%%%%%%%%%%%%%%%%%%%%%%%%%%%%%%%%%%%%%%%%%%%%%%%%%%%%
%%%%%%%%%%%%%%%%%%%%%%%%%%%%%%%%%%%%%%%%%%%%%%%%%%%%%%%%%%%%%%%%%%%%%%%%%%%%%%%%
%%%%%%%%%%%%%%%%%%%%%%%%%%%%%%%%%%%%%%%%%%%%%%%%%%%%%%%%%%%%%%%%%%%%%%%%%%%%%%%%
%%%%%%%%%%%%%%%%%%%%%%%%%%%%%%%%%%%%%%%%%%%%%%%%%%%%%%%%%%%%%%%%%%%%%%%%%%%%%%%%
%%%%%%%%%%%%%%%%%%%%%%%%%%%%%%%%%%%%%%%%%%%%%%%%%%%%%%%%%%%%%%%%%%%%%%%%%%%%%%%%
%%%%%%%%%%%%%%%		Supplementary Material		%%%%%%%%%%%%%%%%%%%%%%%%
%%%%%%%%%%%%%%%		Supplementary Material		%%%%%%%%%%%%%%%%%%%%%%%%
%%%%%%%%%%%%%%%		Supplementary Material		%%%%%%%%%%%%%%%%%%%%%%%%
%%%%%%%%%%%%%%%%%%%%%%%%%%%%%%%%%%%%%%%%%%%%%%%%%%%%%%%%%%%%%%%%%%%%%%%%%%%%%%%%
%%%%%%%%%%%%%%%%%%%%%%%%%%%%%%%%%%%%%%%%%%%%%%%%%%%%%%%%%%%%%%%%%%%%%%%%%%%%%%%%
%%%%%%%%%%%%%%%%%%%%%%%%%%%%%%%%%%%%%%%%%%%%%%%%%%%%%%%%%%%%%%%%%%%%%%%%%%%%%%%%
%%%%%%%%%%%%%%%%%%%%%%%%%%%%%%%%%%%%%%%%%%%%%%%%%%%%%%%%%%%%%%%%%%%%%%%%%%%%%%%%
%%%%%%%%%%%%%%%%%%%%%%%%%%%%%%%%%%%%%%%%%%%%%%%%%%%%%%%%%%%%%%%%%%%%%%%%%%%%%%%%
%%%%%%%%%%%%%%%%%%%%%%%%%%%%%%%%%%%%%%%%%%%%%%%%%%%%%%%%%%%%%%%%%%%%%%%%%%%%%%%%
%%%%%%%%%%%%%%%%%%%%%%%%%%%%%%%%%%%%%%%%%%%%%%%%%%%%%%%%%%%%%%%%%%%%%%%%%%%%%%%%

\newpage 
\clearpage
\onecolumngrid

% ------------------------------------------------------
\begin{center}
	{\large
	Power-law Temperature Dependence of the Penetration Depth in a Topological Superconductor due to Surface States
	\vspace{4pt}
	\\
	SUPPLEMENTAL MATERIAL
	}
\end{center}
% ------------------------------------------------------
%%%%%%%%%% Prefix a "S" to all equations, figures, tables and reset the counter %%%%%%%%%%
\counterwithin{figure}{section}
\counterwithin{equation}{section}
\makeatletter
\setcounter{equation}{0}
\setcounter{figure}{0}
\setcounter{table}{0}
\setcounter{page}{1}
\renewcommand{\theequation}{S\arabic{equation}}
\renewcommand{\thefigure}{S\arabic{figure}}
\renewcommand{\bibnumfmt}[1]{[S#1]}
\renewcommand{\citenumfont}[1]{S#1}
%%%%%%%%%% Prefix a "S" to all equations, figures, tables 

\begingroup
\hypersetup{linkbordercolor=white}
\tableofcontents
\endgroup

\section{I.\ Majorana surface states }

We solve for the surface states by converting $k_z \rightarrow -i\partial_z$ in Eq.~(6),
\begin{equation}\label{eq:surf_state_ode}
	\hat{h}(\textbf{k},-i\partial_z) 
	\,
	\psi_{\varepsilon}(z) 
	= 
	\varepsilon
	\,
	\psi_{\varepsilon}(z),
\end{equation}
where $\vex{k} = (k_x,k_y)$ specifies the momentum transverse to the vacuum-TSC interface at 
$z = 0$ (Fig.~1). 
For this model with a hard wall boundary condition at $z = 0$, 
the surface Majorana fluid has an exactly linear dispersion relation 
$\varepsilon = \Delta k$, corresponding to the surface wave function [Eq.~(8)]
\begin{equation}\label{psiSExp}
	\psi_\textbf{k}^{\mathsf{s}}(z) 
	= 
	\frac{1}{\sqrt{N_\textbf{k}^{\mathsf{s}}}} 
	\,
	e^{-z/l_{\mathsf{coh}}}
	\,
	\sin\left(z\sqrt{\lambdabar_F^{-2}-l_{\mathsf{coh}}^{-2}-k^2} \right)
	\begin{bmatrix}
	1				\\
	-ie^{i\phi_\textbf{k}}		\\
	1				\\
	i e^{i\phi_\textbf{k}}
	\end{bmatrix},
\end{equation} 
where 
$
	e^{i\phi_\textbf{k}} \equiv (k_x + ik_y)/k
$ 
and the normalization constant 
\begin{align}
	N_{\vex{k}}^{\mathsf{s}} 
	= 
	l_{\mathsf{coh}}
	\frac{
	\left(
		\lambdabar_F^{-2} - l_{\mathsf{coh}}^{-2} - k^2  
	\right)
	}{
	\left(
		\lambdabar_F^{-2} - k^2
	\right)
	}. 
\end{align}
The four-component spinor in Eq.~(\ref{psiSExp}) is expressed in the [spin ($s$)]$\otimes$[particle-hole ($\sigma$)] basis
such that $\hat{s}^3 \rightarrow \diag(1,-1,1,-1)$ and $\hat{\sigma}^3 \rightarrow \diag(1,1,-1,-1)$. 

The Bogoliubov-de Gennes Hamiltonian $\hat{h}(\vex{k})$ in Eq.~(6) has the following particle-hole (P), 
time-reversal (T), 
and chiral (S) symmetries:
\begin{eqnarray}
	-
	\hat{M}_{\mathsf{P}}^{-1}
	\,
	\hat{h}^{\mathsf{T}}(-\textbf{k})
	\,
	\hat{M}_{\mathsf{P}} 
	&=& 
	\hat{h}(\textbf{k}),
	\qquad  
	\hat{M}_{\mathsf{P}} = \hat{s}^2 \hat{\sigma}^2 = \hat{M}_{\mathsf{P}}^{\mathsf{T}},
\\
	\hat{M}_{\mathsf{T}}^{-1}
	\,
	\hat{h}^{*}(-\textbf{k}) 
	\,
	\hat{M}_{\mathsf{T}} 
	&=& 
	\hat{h}(\textbf{k}),
	\qquad  
	\hat{M}_{\mathsf{T}} = i\hat{s}^2 \hat{\sigma}^3 = -\hat{M}_{\mathsf{T}}^{\mathsf{T}},
\\
	-
	\hat{M}_{\mathsf{S}}
	\,
	\hat{h}(\textbf{k})
	\,
	\hat{M}_{\mathsf{S}} 
	&=& 
	\hat{h}(\textbf{k}),
	\qquad  
	\hat{M}_{\mathsf{S}} = \hat{\sigma}^1, 
\end{eqnarray}
where $\hat{A}^{\mathsf{T}}$ is the matrix transpose of $\hat{A}$. 
The chiral symmetry is a product of time-reversal and particle-hole;
since the latter is an automatic consequence of fermion antisymmetry, 
chiral is equivalent to time-reversal. 

The negative-energy surface eigenstate with momentum $\vex{k}$ is the chiral transform of
Eq.~(\ref{psiSExp}), $\hat{\sigma}^1 \, \psi_\textbf{k}^{\mathsf{s}}(z)$.
Positive- and negative-energy surface states are \emph{bi-locally} orthogonal
(due to the spin-momentum--locked spinors), 
\begin{equation}\label{Ortho}
	\psi_\textbf{k}^{\mathsf{s}\dagger}(z_1) \, \hat{\sigma}^1 \, \psi_\textbf{k}^{\mathsf{s}}(z_2) 
	= 
	\psi_\textbf{k}^{\mathsf{s}\dagger}(z_1) \, \hat{s}^3 \, \psi_\textbf{k}^{\mathsf{s}}(z_2) 
	= 	
	0.
\end{equation}

%%%%%%%%%%%%%%%%%%%%%%%%%%%%%%%%%%%%%%%%%%%%%%%%%%%%%%%%%%%%%%%%%%%%%%%%%%%%%%%%
%%%%%%%%%%%%%%%%%%%%%%%%%%%%%%%%%%%%%%%%%%%%%%%%%%%%%%%%%%%%%%%%%%%%%%%%%%%%%%%%
%%%%%%%%%%%%%%%%%%%%%%%%%%%%%%%%%%%%%%%%%%%%%%%%%%%%%%%%%%%%%%%%%%%%%%%%%%%%%%%%
%%%%%%%%%%%%%%%%%%%%%%%%%%%%%%%%%%%%%%%%%%%%%%%%%%%%%%%%%%%%%%%%%%%%%%%%%%%%%%%%
%%%%%%%%%%%%%%%%%%%%%%%%%%%%%%%%%%%%%%%%%%%%%%%%%%%%%%%%%%%%%%%%%%%%%%%%%%%%%%%%
%%%%%%%%%%%%%%%%%%%%%%%%%%%%%%%%%%%%%%%%%%%%%%%%%%%%%%%%%%%%%%%%%%%%%%%%%%%%%%%%
%%%%%%%%%%%%%%%%%%%%%%%%%%%%%%%%%%%%%%%%%%%%%%%%%%%%%%%%%%%%%%%%%%%%%%%%%%%%%%%%
%%%%%%%%%%%%%%%%%%%%%%%%%%%%%%%%%%%%%%%%%%%%%%%%%%%%%%%%%%%%%%%%%%%%%%%%%%%%%%%%
%%%%%%%%%%%%%%%%%%%%%%%%%%%%%%%%%%%%%%%%%%%%%%%%%%%%%%%%%%%%%%%%%%%%%%%%%%%%%%%%
%%%%%%%%%%%%%%%%%%%%%%%%%%%%%%%%%%%%%%%%%%%%%%%%%%%%%%%%%%%%%%%%%%%%%%%%%%%%%%%%
%%%%%%%%%%%%%%%%%%%%%%%%%%%%%%%%%%%%%%%%%%%%%%%%%%%%%%%%%%%%%%%%%%%%%%%%%%%%%%%%
%%%%%%%%%%%%%%%%%%%%%%%%%%%%%%%%%%%%%%%%%%%%%%%%%%%%%%%%%%%%%%%%%%%%%%%%%%%%%%%%
%%%%%%%%%%%%%%%%%%%%%%%%%%%%%%%%%%%%%%%%%%%%%%%%%%%%%%%%%%%%%%%%%%%%%%%%%%%%%%%%
%%%%%%%%%%%%%%%%%%%%%%%%%%%%%%%%%%%%%%%%%%%%%%%%%%%%%%%%%%%%%%%%%%%%%%%%%%%%%%%%

\section{II.\ Paramagnetic Current-Current Correlation Function}

\subsection{A.\ Correlation function from the bulk alone}

The imaginary time action corresponding to Eq.~(5) is
\begin{eqnarray}
	S 
	&=& 
	\frac{T}{2}
	\sum_{\omega_n} 
	\int 
	d^3\vex{r}
	\,
	\chi^{\mathsf{T}}(-\omega_n,\textbf{r}) 
	\,
	i\hat{M}_{\mathsf{P}}
	\left(
		-i \omega_n + \hat{h}
	\right)
	\chi(\omega_n,\textbf{r}),
\end{eqnarray}
where $\omega_n$ denotes a fermionic Matsubara frequency. 
The imaginary time paramagnetic current-current correlation function is
\begin{equation}\label{eq:Pi_bulk}
\begin{aligned}
	\Pi_1^{xx}(\tau,\textbf{r},\textbf{r}')
	=&\,
	\left\langle J_1^x(\tau,\textbf{r}) \, J_1^{x}(0,\textbf{r}')\right\rangle
\\
	=&\,
	-
	\frac{1}{2}
	\left(\frac{e}{m}\right)^2
	T^2
	\sum_{\omega_1,\omega_2}
	\int_{\textbf{k}_1,\textbf{k}_2}	
	e^{-i(\omega_1-\omega_2)\tau+i(\textbf{k}_1-\textbf{k}_2)\cdot(\textbf{r}-\textbf{r}')}
	\,
	(k_1^x)(k_2^x)
	\,
	\mathsf{Tr}\left[
		\hat{G}(i\omega_1,\textbf{k}_1)	\,\hat{G}(i\omega_2,\textbf{k}_2)
	\right],
\end{aligned}
\end{equation}
where we have made use of the translational invariance in the bulk. 
Using the Green's function 
$
	\hat{G}(i\omega,\textbf{k}) 
	= 
	\left[-i \omega + \hat{h}(\textbf{k})\right]^{-1},
$ 
the Fourier transformed correlation function is
\begin{equation}\label{eq:Pi_bulk_k}
\begin{aligned}
	\tilde{\Pi}^{xx}_1(i\Omega_n=0,\textbf{q}=0) 
	=&\, 
	-
	\frac{2}{3}
	\left(\frac{e}{m}\right)^2 
	T
	\sum_{\omega}
	\int_\textbf{k} 
	k^2 
	\,
	\frac{E_k^2-\omega^2}{(\omega^2+E_k^2)^2}
\\
	=&\, 
	\frac{\beta}{6}
	\left(\frac{e}{m}\right)^2 
	\int_\textbf{k} 
	k^2 
	\sech^2\left( \frac{\beta E_k}{2}\right),
\end{aligned}
\end{equation}
where $E_k = \sqrt{\tilde{\varepsilon}_k^2 + \Delta^2 k^2}$. The retarded version is given by Eq.~(7) in the main text.

\subsection{B.\ Correlation functions for the semi-infinite slab}

To consider the effect of the surface, we must retain the $(z,z')$-dependence of the Green's function. 
Eqs. (\ref{eq:Pi_bulk}) and (\ref{eq:Pi_bulk_k}) are replaced by
\begin{equation}
	\Pi_1^{xx}(\tau,\textbf{r}-\textbf{r}',z,z') 
	= 
	-
	\frac{1}{2}
	\left(\frac{e}{m}\right)^2 T^2
	\sum_{\omega_1,\omega_2} 
	\int_{\textbf{k}_1,\textbf{k}_2}
	e^{-i(\omega_1-\omega_2)\tau+i(\textbf{k}_1-\textbf{k}_2)\cdot (\textbf{r}-\textbf{r}')} 
	(k_1^x)(k_2^x)
	\,
	\mathsf{Tr}
	\left[
		\hat{G}(i\omega_1,\textbf{k}_1;z,z')
		\,
		\hat{G}(i\omega_2,\textbf{k}_2;z',z)
	\right]\!,
\end{equation}
where $\textbf{r} = (x,y)$ and $\textbf{k} = (k_x,k_y)$ are 2D vectors parallel to the interface. 
Then
\begin{equation}
	\tilde{\Pi}_1^{xx}(i\Omega_n = 0, \textbf{q}=0;z,z')
	=
	-
	\frac{1}{2} 
	\left(\frac{e}{m}\right)^2 
	T \sum_{\omega}
	\int_\textbf{k} 
	(k^x)^2 
	\,
	\mathsf{Tr}
	\left[
		\hat{G}(i\omega_1,\textbf{k};z,z')
		\,
		\hat{G}(i\omega_2,\textbf{k};z',z)
	\right].
\end{equation}
We assume a generic eigenstate decomposition for $\hat{G}$,
\begin{equation}
	\hat{G}(i\omega,\textbf{k};z,z') 
	= 
	\sum_{\varepsilon} 
	\frac{
	\psi_{\varepsilon,\textbf{k}}(z) \, \psi^{\dagger}_{\varepsilon,\textbf{k}}(z')
	}{
	-i\omega +\varepsilon
	},
\end{equation}
where the sum runs over all positive- and negative-energy bulk and surface states of $\hat{h}$ in 
Eq.~(\ref{eq:surf_state_ode}),
so that
\begin{equation}\label{PixxIT}
	\tilde{\Pi}_1^{xx}(i\Omega_n=0,\textbf{q}=0;z,z')
	=
	\frac{1}{2}
	\left(\frac{e}{m}\right)^2 
	\int_\textbf{k} 
	(k^x)^2
	\sum_{\varepsilon_1,\varepsilon_2}
	\left[
		\psi_{\varepsilon_2,\textbf{k}}^{\dagger}(z) 
		\,
		\psi_{\varepsilon_1,\textbf{k}}(z) 
		\,
		\psi_{\varepsilon_1,\textbf{k}}^{\dagger}(z')
		\, 
	\psi_{\varepsilon_2,\textbf{k}}(z') 
	\right]
	F(\varepsilon_1,\varepsilon_2),
\end{equation}
where
\begin{align}
	F(\varepsilon_1,\varepsilon_2)
	\equiv&\,
	-
	T
	\sum_{\omega}
	\frac{1}{(-i\omega+\varepsilon_1)(-i\omega+\varepsilon_2)}
	=
	\left\{
	\begin{aligned}
		&\,
		\displaystyle{
		\frac{
			\tanh\left(\frac{\beta \varepsilon_1}{2}\right) 
			-  
			\tanh\left(\frac{\beta \varepsilon_2}{2}\right)
		}{
			2(\varepsilon_1-\varepsilon_2)
		},
		}
		\qquad 
		&
		\varepsilon_1 \neq \varepsilon_2,
	\\
		&\,
		\frac{\beta}{4} \sech^2\left(\frac{\beta \varepsilon_1}{2}\right), 
		&
		\varepsilon_1 = \varepsilon_2.
		\end{aligned}
		\right.
\end{align}
For a system that is isotropic (rotationally invariant) in the $(x,y)$ plane parallel to the 
interface, 
the double-Fourier transform of the retarded version is
\begin{align}\label{eq:double_FT}
	\tilde{\Pi}_{1,R}^{xx}(0,0,0;q_z,-Q_z)
	=
	-
	\frac{1}{4}
	\left(\frac{e}{m}\right)^2
	\int_\textbf{k}k^2
	\sum_{\varepsilon_1,\varepsilon_2}
	F(\varepsilon_1,\varepsilon_2)
	&\,
	\left[
		\int_0^{\infty}dz	
		\,
		e^{-i q_z z}
		\,
		\psi_{\varepsilon_2,\textbf{k}}^{\dagger}(z) 
		\,
		\psi_{\varepsilon_1,\textbf{k}}(z) 		
	\right]
\nonumber\\
	\times
	&\,
	\left[
		\int_0^{\infty}dz'	
		\,
		e^{i Q_z z'}
		\,
		\psi_{\varepsilon_1,\textbf{k}}^{\dagger}(z')
		\, 
		\psi_{\varepsilon_2,\textbf{k}}(z') 
	\right].
\end{align}

\subsubsection{1.\ Surface-surface response}

The surface eigenstates 
$\psi_{\vex{k}}^{\mathsf{s}}(z)$
and 
$\hat{\sigma}^1 \, \psi_{\vex{k}}^{\mathsf{s}}(z)$
[Eqs.~(8) and (\ref{psiSExp})]
respectively 
have 
eigenenergies 
$\pm \Delta |\vex{k}|$.  
The surface-surface contribution to Eq.~(\ref{eq:double_FT}) is
\begin{equation}\label{SSPi}
\begin{aligned}[b]
	\tilde{\Pi}_{1,R;\sfs,\sfs}^{xx} 
%	=&\,
%	-
%	\frac{1}{2}
%	\left(\frac{e}{m}\right)^2
%	\int_\textbf{k}k^2 
%	\int_0^{\infty} dz \int_0^{\infty}dz'
%	\,
%	e^{-iq_z z + iQ_z z'}
%	\left[
%	\begin{aligned}
%	&\,
%		F(\Delta k,\Delta k) \,
%		\psi_\textbf{k}^{\mathsf{s}\dagger}(z)
%		\,
%		\psi_\textbf{k}^{\mathsf{s}}(z)
%		\,
%		\psi_\textbf{k}^{\mathsf{s}\dagger}(z')
%		\,
%		\psi_\textbf{k}^{\mathsf{s}}(z')
%	\\
%	&\,
%		+
%		F(\Delta k,-\Delta k) \,
%		\psi_\textbf{k}^{\mathsf{s}\dagger}(z)
%		\,
%		\hat{\sigma}^1
%		\,
%		\psi_\textbf{k}^{\mathsf{s}}(z)
%		\,
%		\psi_\textbf{k}^{\mathsf{s}\dagger}(z')
%		\,
%		\hat{\sigma}^1
%		\,
%		\psi_\textbf{k}^{\mathsf{s}}(z')
%	\end{aligned}
%	\right]
%\\
	=&\,
	-
	\frac{1}{2}
	\left(\frac{e}{m}\right)^2
	\int_\textbf{k} 
	\frac{\beta k^2}{4} \sech^2\left(\frac{\beta \Delta k}{2}\right)
	\left[
		\int_0^{\infty}dz
		\,
		e^{-iq_z z} 
		\psi_\textbf{k}^{\mathsf{s}\dagger}(z)
		\,
		\psi_\textbf{k}^{\mathsf{s}}(z)
	\right]
	\left[
		\int_0^{\infty}dz'
		\,
		e^{iQ_z z'} 
		\,
		\psi_\textbf{k}^{\mathsf{s}\dagger}(z')
		\,
		\psi_\textbf{k}^{\mathsf{s}}(z')
	\right]
\\
	=&\,
	-
	\frac{\beta}{2^4 \pi}
	\left(\frac{e}{m}\right)^2
	\int_0^{\lambdabar_F^{-1}}dk\, k^3 
	\sech^2\left(\frac{\beta \Delta k}{2}\right)
	\,
	\Theta(k,q_z)
	\,
	\Theta(k,-Q_z),
\end{aligned}
\end{equation}
where we have used Eq.~(\ref{Ortho}), and where
[Eq.~(12)]
\begin{equation}
	\Theta(k,q_z)
	\equiv
	\int_0^{\infty}dz
	\,
	e^{-iq_z z}
	\,
	\psi_\textbf{k}^{\mathsf{s}\dagger}(z)
	\,
	\psi_\textbf{k}^{\mathsf{s}}(z)
	=
	\frac{
	8 i l_{\mathsf{coh}}^{-1} \left(k^2 - \lambdabar_F^{-2}\right)
	}{
	\left[2 i l_{\mathsf{coh}}^{-1} - q_z\right]
	\left[
	4\left(k^2 - \lambdabar_F^{-2}\right)
	-
	4 i l_{\mathsf{coh}}^{-1} q_z 
	+ 
	q_z^2
	\right]
	}.
\end{equation}
The ultraviolet momentum cutoff $\lambdabar_F^{-1} = k_F$ in Eq.~(\ref{SSPi})
is where the surface Majorana band merges with the bulk quasiparticle continuum. 
For low temperatures $\beta\rightarrow\infty$, we can extend the upper limit 
of the $k$ integration to infinity, and drop the dependence of $\Theta(k,q)$ on 
$k$. To leading order in temperature, one obtains Eq.~(11) in the main text.

\subsubsection{2.\ Surface-bulk cross terms}

The surface-bulk cross term contributions to Eq.~(\ref{eq:double_FT}) take the form
\begin{equation}
	\tilde{\Pi}_{1,R,\sfs,\sfb}^{xx}(0,0,0;q_z,-Q_z) 
	=
	-
	\left(\frac{e}{m}\right)^2
	\int_0^{\lambdabar_F^{-1}}dk\, k^3  
	\sum_{\sfs,\sfb} 
	\alpha_{\sfs,\sfb}(\vex{k};q_z,-Q_z) 
	\,
	F\left[E_{\sfs}(\vex{k}),E_{\sfb}(\vex{k})\right],  
\end{equation}
where $\alpha_{\sfs,\sfb}(\vex{k};q_z,-Q_z)$ is a temperature-independent 
coefficient encoding the $(q_z,-Q_z)$-transformed overlaps between the surface bound Majorana and bulk standing wave quasiparticle states. The summation runs over all surface and bulk eigenstates with eigenenergies $E_{\sfs}(\vex{k})$ and $E_{\sfb}(\vex{k})$, respectively. 
At low temperatures, the expression is dominated by small energies, but the mismatch between 
the gapless surface 
[$\lim_{k \rightarrow 0} |E_{\sfs}(\vex{k})| = \Delta k \rightarrow 0$]
and 
the gapped bulk
[$\min|E_{\sfb}(\vex{k})| = \Delta k_F$]
means that the leading temperature dependence of this term is 
\begin{equation}
	\tilde{\Pi}_{1,R,\sfs,\sfb}^{xx}(0,0,0;q_z,-Q_z) 
	\sim
	T^4,
\end{equation}
which is higher order than the contribution of the surface-surface term.

%%%%%%%%%%%%%%%%%%%%%%%%%%%%%%%%%%%%%%%%%%%%%%%%%%%%%%%%%%%%%%%%%%%%%%%%%%%%%%%%
%%%%%%%%%%%%%%%%%%%%%%%%%%%%%%%%%%%%%%%%%%%%%%%%%%%%%%%%%%%%%%%%%%%%%%%%%%%%%%%%
%%%%%%%%%%%%%%%%%%%%%%%%%%%%%%%%%%%%%%%%%%%%%%%%%%%%%%%%%%%%%%%%%%%%%%%%%%%%%%%%
%%%%%%%%%%%%%%%%%%%%%%%%%%%%%%%%%%%%%%%%%%%%%%%%%%%%%%%%%%%%%%%%%%%%%%%%%%%%%%%%
%%%%%%%%%%%%%%%%%%%%%%%%%%%%%%%%%%%%%%%%%%%%%%%%%%%%%%%%%%%%%%%%%%%%%%%%%%%%%%%%
%%%%%%%%%%%%%%%%%%%%%%%%%%%%%%%%%%%%%%%%%%%%%%%%%%%%%%%%%%%%%%%%%%%%%%%%%%%%%%%%
%%%%%%%%%%%%%%%%%%%%%%%%%%%%%%%%%%%%%%%%%%%%%%%%%%%%%%%%%%%%%%%%%%%%%%%%%%%%%%%%
%%%%%%%%%%%%%%%%%%%%%%%%%%%%%%%%%%%%%%%%%%%%%%%%%%%%%%%%%%%%%%%%%%%%%%%%%%%%%%%%
%%%%%%%%%%%%%%%%%%%%%%%%%%%%%%%%%%%%%%%%%%%%%%%%%%%%%%%%%%%%%%%%%%%%%%%%%%%%%%%%
%%%%%%%%%%%%%%%%%%%%%%%%%%%%%%%%%%%%%%%%%%%%%%%%%%%%%%%%%%%%%%%%%%%%%%%%%%%%%%%%
%%%%%%%%%%%%%%%%%%%%%%%%%%%%%%%%%%%%%%%%%%%%%%%%%%%%%%%%%%%%%%%%%%%%%%%%%%%%%%%%
%%%%%%%%%%%%%%%%%%%%%%%%%%%%%%%%%%%%%%%%%%%%%%%%%%%%%%%%%%%%%%%%%%%%%%%%%%%%%%%%
%%%%%%%%%%%%%%%%%%%%%%%%%%%%%%%%%%%%%%%%%%%%%%%%%%%%%%%%%%%%%%%%%%%%%%%%%%%%%%%%
%%%%%%%%%%%%%%%%%%%%%%%%%%%%%%%%%%%%%%%%%%%%%%%%%%%%%%%%%%%%%%%%%%%%%%%%%%%%%%%%

\section{III.\ Low-temperature field penetration}

\subsection{A.\ Solution to the integral equation}

The kernel in Eq.~(11) can be identified as the matrix elements of an outer product
\begin{equation}
	\tilde{\Pi}_{1,R;\sfs,\sfs}^{xx} 
	= 
	-
	\Upsilon 
	\,
	\braket{q_z | R} \braket{R | Q_z},
\qquad
	\Upsilon
	\equiv
	\left[\frac{2^3 3^2 \zeta(3)}{\pi}\right] 
	\left(\frac{2 m \mu e}{\Delta}\right)^2
	(k_B T)^3.
\end{equation}
Here we assume the norm and resolution of the identity,
\begin{equation}
	\braket{q_z|Q_z} = 2\pi \delta (q_z -Q_z),
	\qquad \qquad
	\hat{1} = \int_{-\infty}^{\infty} \frac{dq_z}{2\pi} \ket{q_z}\bra{q_z}.
\end{equation}
The Meissner response in Eq.~(3) can then be written as
\begin{equation}
\begin{aligned}[b]
	-
	2 B_0 
	\int_{Q_z} 
	\braket{q_z|Q_z}
	=&\,
	\left[q_z^2 + \left(\LL\right)^{-2}\right]\tilde{A}(q_z)
	-
	\frac{4\pi\Upsilon}{c^2} 
	\int_{Q_z} 
	\braket{q_z|R}\braket{R|Q_z}\tilde{A}(Q_z)
\\
	=&\,
	\bra{q_z} 
	\left[
		\hat{M}(\hat{q}_z)- \frac{4\pi \Upsilon}{c^2}\ket{R}\bra{R}
	\right]
	\ket{\tilde{A}},
	\qquad
	\hat{M}(\hat{q}_z) 
	\equiv 
	\left[ 
		\hat{q}_z^2 + \left(\LL\right)^{-2}
	\right].
\end{aligned}
\end{equation}
Formally, we can invert the operator to obtain
\begin{equation}\label{AFTSol}
\begin{aligned}[b]
	\tilde{A}(q_z) 
	=&\,
	-2 B_0
	\bra{q_z}
	\int_{Q_z}
	\hat{M}^{-1}(\hat{q}_z)
	\left[
		\hat{1} - \frac{4\pi \Upsilon}{c^2}\ket{R}\bra{R} \hat{M}^{-1}(\hat{q}_z)
	\right]^{-1}
	\ket{Q_z}
\\
	=&\,
 	-
 	2 B_0
 	\left\{ 
 		M^{-1}(q_z) 
 		+
  		\left(\frac{4\pi \Upsilon \Xi}{c^2}\right) 
  		M^{-1}(q_z) 
  		\,
  		R(q_z)
  		\int_{Q_z}
  		M^{-1}(Q_z) 
  		\,
  		R^*(Q_z)
	\right\},
\end{aligned}
\end{equation}
where
\begin{equation}
	\Xi(T) 
	\equiv 
	\left\{
		1	
		-  
		\left(\frac{4\pi \Upsilon}{c^2}\right)
		\left[\bra{R}\hat{M}^{-1}(\hat{q}_z)\ket{R}\right]
	\right\}^{-1}
\end{equation}
is a temperature-dependent constant that goes to $1 + \ord{T^3}$ as $k_B T \rightarrow 0$. 
Using 
$l_{\mathsf{coh}} = (m\Delta)^{-1}$, 
$\lambdabar_F = 1/\sqrt{2m\mu}$, 
$\LL = \sqrt{m c^2/(4 \pi e^2 n)}$,
$n = k_F^3/3\pi^2$, 
and defining physical BCS gap as $\Delta_{BCS} = k_F \Delta$, 
we define  
\begin{equation}
	\varrho(T) 
	\equiv
	\frac{8\pi \Upsilon}{c^2} \left(\LL\right)^8 
	= 
	\bigg[2^4 3^3 \zeta(3) \pi  \Xi(T)\bigg]
	\frac{ \left(\LL\right)^6}{l_{\mathsf{coh}}\lambdabar_F^4} 
	\left(\frac{k_B T}{\Delta_{BCS}}\right)^3,
\end{equation} 
which is the length scale introduced in Eq.~(14), 
after replacing $\Xi(T) \rightarrow 1$ (valid in the low-temperature limit). 
Then Eq.~(\ref{AFTSol}) becomes
\begin{equation}\label{ASol}
	A(z) 
	= 
	-
	B_0 
	\left\{
		\LL
		\,
		e^{-z/\LL} 
		+
		\varrho(T) 
		\,
		G(z)
		\,
		G(0)
	\right\},
\end{equation}
where 
the dimensionless function $G(z)$ is given by
\begin{equation}
	G(z) 
	\equiv
	\frac{1}{ \left(\LL\right)^{4}}
	\int_{q_z}
	e^{i q_z z}
	\,
	M^{-1}(q_z)
	\,
	R(q_z) 
	=
  	\frac{i}{ \left(\LL\right)^{4}}
  	\int_{q_z}
 	\frac{e^{iq_z z}}{
	\left[q_z^2 + \left(\LL\right)^{-2}\right]
	\left[q_z - 2 i l_{\mathsf{coh}}^{-1}\right]
	\left[q_z^2 - 4 i l_{\mathsf{coh}}^{-1} q_z - 4 \lambdabar_F^{-2}\right]
}.
\end{equation}
In the weak-pairing BCS limit ($\lambdabar_F < l_{\mathsf{coh}}$),
this evaluates to 
\begin{align}\label{G(z)Eval}
	G(z)
	=&\,
	-
	\frac{
	e^{- z / \LL}
	}{
	2 \left(\LL\right)^{3}
	\left[\left(\LL\right)^{-1} - 2 \lcoh^{-1}\right]
	\left[\left(\LL\right)^{-2} + 4 \lfb^{-2} - 4 \lcoh^{-1} \left(\LL\right)^{-1}\right]	
	}
\nonumber\\
\phantom{0}
\nonumber\\
&\,
+	
	\frac{
	1
	}{
	\left(\LL\right)^{4}
	\left[
		\left(\LL\right)^{-2} - 4 \lcoh^{-2}
	\right]
	}
	\frac{
	e^{- 2 z / \lcoh}
	}{
	4
	(\lfb^{-2} - \lcoh^{-2})
	}
\nonumber\\
\phantom{0}
\nonumber\\
&\,
+
\left[
\begin{aligned}
	\left[
		\lfb^2 \left(\LL\right)^{-2} 
		\left\{
			\lcoh^2 \left[\lfb^2+4 \left(\LL\right)^2\right]
			-
			8 \lfb^2 \left[\LL\right]^2
		\right\}
	\right]
	&\,
	\cos\left(2 \sqrt{\lfb^{-2} - \lcoh^{-2}} \, z\right) 
\\
+
	\left[
		8 
		\lfb^4 \lcoh 
		\sqrt{\lfb^{-2} - \lcoh^{-2}} 
	\right]
	&\,	
	\sin\left(2 \sqrt{\lfb^{-2} - \lcoh^{-2}} \, z\right)
\end{aligned}
\right]
\nonumber\\
&\,
\phantom{+}\,\,
\times
	\frac{
	e^{- 2 z / \lcoh}
	}{
	4
	(\lfb^{-2} - \lcoh^{-2})
	\left\{
		16 \lfb^4 
		\left(\LL\right)^2
		-
		\lcoh^2 
		\left[	
			\lfb^2+4 \left(\LL\right)^2
		\right]^2
	\right\}
	}.
\end{align}

From Eq.~(\ref{ASol}), the magnetic field inside the slab is given by Eq.~(13) in the main text. 
The results in Eqs.~(16) and (17) obtain from the 
type I 
($\{\lfb,\LL\} \ll \lcoh$) 
and 
type II
($\lfb \ll \lcoh \ll \LL$) 
limits of Eq.~(\ref{G(z)Eval}).

\subsection{B.\ Penetration Depth}

In the expression for the effective penetration depth given by Eq.~(15), 
the parameter $G(0)$ evaluates to [Eq.~(\ref{G(z)Eval})]
\begin{equation}\label{G(0)Eval}
	G(0) 
	= 
	\frac{
	1
}{
	2
	\left(1+2 \LL l_{\mathsf{coh}}^{-1}\right)
	\left[
		1 + 4\LL \left(l_{\mathsf{coh}}^{-1} + \LL \lambdabar_F^{-2}\right)
	\right]
}.
\end{equation}
In the type-I limit ($\{\lfb,\LL\} \ll \lcoh$), 
this simplifies to 
\begin{equation}
	G_{\mathrm{I}}(0) 
	\simeq
	\frac{
	\lambdabar_F^2
	}{
	2\left[\lambdabar_F^2 + 4\left(\LL\right)^2\right]
	}.
\end{equation}
In the opposite type-II limit ($\{\lfb,\lcoh\} \ll \LL$),
Eq.~(\ref{G(0)Eval}) instead becomes 
\begin{equation}
	G_{\mathrm{II}}(0)  
	\simeq
	\frac{l_{\mathsf{coh}} \lambdabar_F^2}{16 \left(\LL\right)^{3}}.
\end{equation}

\end{document}